\documentclass[aps,twocolumn,prl,amsmath,amssymb,superscriptaddress,floats]{revtex4}
\usepackage{txfonts}
\usepackage[usenames,dvipsnames]{color}

\usepackage{graphicx}
\usepackage{dcolumn}
\usepackage{bm}
\usepackage{hyperref}
\usepackage[normalem]{ulem}
\usepackage{caption}
\usepackage{subcaption}
\newcommand{\upperRomannumeral}[1]{\uppercase\expandafter{\romannumeral#1}}

\newcommand{\red}[1]{{\textcolor{red}{#1}}}

\begin{document}

\title{A new type of cyclotron resonance from charge-impurity scattering in the bulk-insulating Bi$_2$Se$_3$ thin films }



\author{Xingyue Han}
\affiliation{Department of Physics and Astronomy, University of Pennsylvania, Philadelphia, PA 19104 USA.}
\author{Maryam Salehi}
\affiliation{Department of Material Science and Engineering, Rutgers the State University of New Jersey, Piscataway, NJ 08854 USA.}
\author{Seongshik Oh}
\affiliation{Department of Physics and Astronomy, Rutgers the State University of New Jersey, Piscataway, NJ 08854 USA.}
 \author{Liang Wu}
 \email{liangwu@sas.upenn.edu}
 \affiliation{Department of Physics and Astronomy, University of Pennsylvania, Philadelphia, PA 19104 USA.}

\begin{abstract}
This work focuses on the low frequency Drude response of bulk-insulating topological insulator Bi$_2$Se$_3$ films. The frequency and field dependence of the mobility and carrier density are measured simultaneously via time-domain terahertz spectroscopy.  These films are grown on buffer layers, capped by Se, and have been exposed in air for months. Under a magnetic field up to 7 Tesla, we observe prominent cyclotron resonances (CRs). We attribute the sharp CR to two different topological surface states (TSSs) from both surfaces of the films. The CR sharpens at high fields due to an electron-impurity scattering. By using magneto-terahertz spectroscopy, we confirm that these films are bulk-insulating, which paves the way to use intrinsic topological insulators without bulk carriers for applications including topological spintronics and quantum computing. 
\end{abstract}

\date{\today}
\maketitle
\textbf{Introduction} 

Ordered states of matter are usually characterized by Landau's spontaneous symmetry breaking theory. For example, ferromagnets break continuous rotation symmetry and superconductors break gauge symmetry. The newly discovered topological materials are classified by the topological properties of their bulk wavefunction. For example,the three-dimensional  topological insulators (TIs) possess insulating bulk states and conducting surface states. \cite{Hasan-Kane-10,Qi-Zhang-11} .    Intrinsic bulk-insulating TIs are the ideal platforms for low-dissipation spintronics and quantum computing due to the spin-momentum locking on the topological surface states (TSSs) \cite{Hasan-Kane-10,Qi-Zhang-11}. However, most TI materials are slightly doped and thus have a conducting bulk, which is the major obstacle to reach the ideal TI state. To solve the problem, many attempts are made to tune the chemical potential towards the Dirac point. It was a success in graphene to probe many-body interactions with plasmons and phonons \cite{BasovRMP14}, and in Cu doped Bi$_2$Se$_3$ with dominating electron-phonon coupling\cite{WuPRL15}. More advancements in TIs are still lacking .

Bi$_2$Se$_3$ is a typical 3D TI with a single Dirac cone on the surface. However, native grown Bi$_2$Se$_3$ is known with a conducting bulk from defects. One way to suppress the bulk carrier density is the chemical doping method \cite{RenPRB2010, XiongPhysicaE2012}.  However, the bulk carrier densities in these samples are still not negligible and the impurity states are pinned close to the chemical potential. The other way to suppress the bulk carriers is to grow Bi$_2$Se$_3$ on a buffer layer \cite{KoiralaNanoLetters15, WuScience16}. An open question remains on how to keep the sample bulk-insulating after exposure in air. In this work, we investigate Bi$_2$Se$_3$ thin films grown on a buffer layer with Se capping. The TSSs are decoupled via time-domain terahertz spectroscopy with a magnetic field up to 7 tesla. 

Cyclotron resonance (CR) measurements have been shown very useful to study the Dirac fermions and many-body interactions \cite{Crassee10a, JiangPRL07a,JenkinsPRB13}. It is also one of the most precise method to determine the carrier effective mass \cite{KonoReview06}. In previous work, Bi$_2$Se$_3$ films with conducting bulk showed a large Kerr rotation without obvious resonance \cite{ValdesAguilarPRL12}. In a later report, CR emerges in In$_2$Se$_3$ capped Bi$_2$Se$_3$ films \cite{JenkinsPRB13} due to the destroyed non-TI/TI interface from Indium diffusion\cite{LeeThinFilms14}. This is because a phase transition from topological to trivial insulator occurs at a low Indium concentration ($\sim 6\%$) \cite{WuNatPhys13}. Besides, Cu doped Bi$_2$Se$_3$ films also presents CR, where the bulk-insulating state is reached and the electron-phonon coupling was revealed \cite{WuPRL15}.


\begin{figure}[htp]
\includegraphics[width=0.5\textwidth]{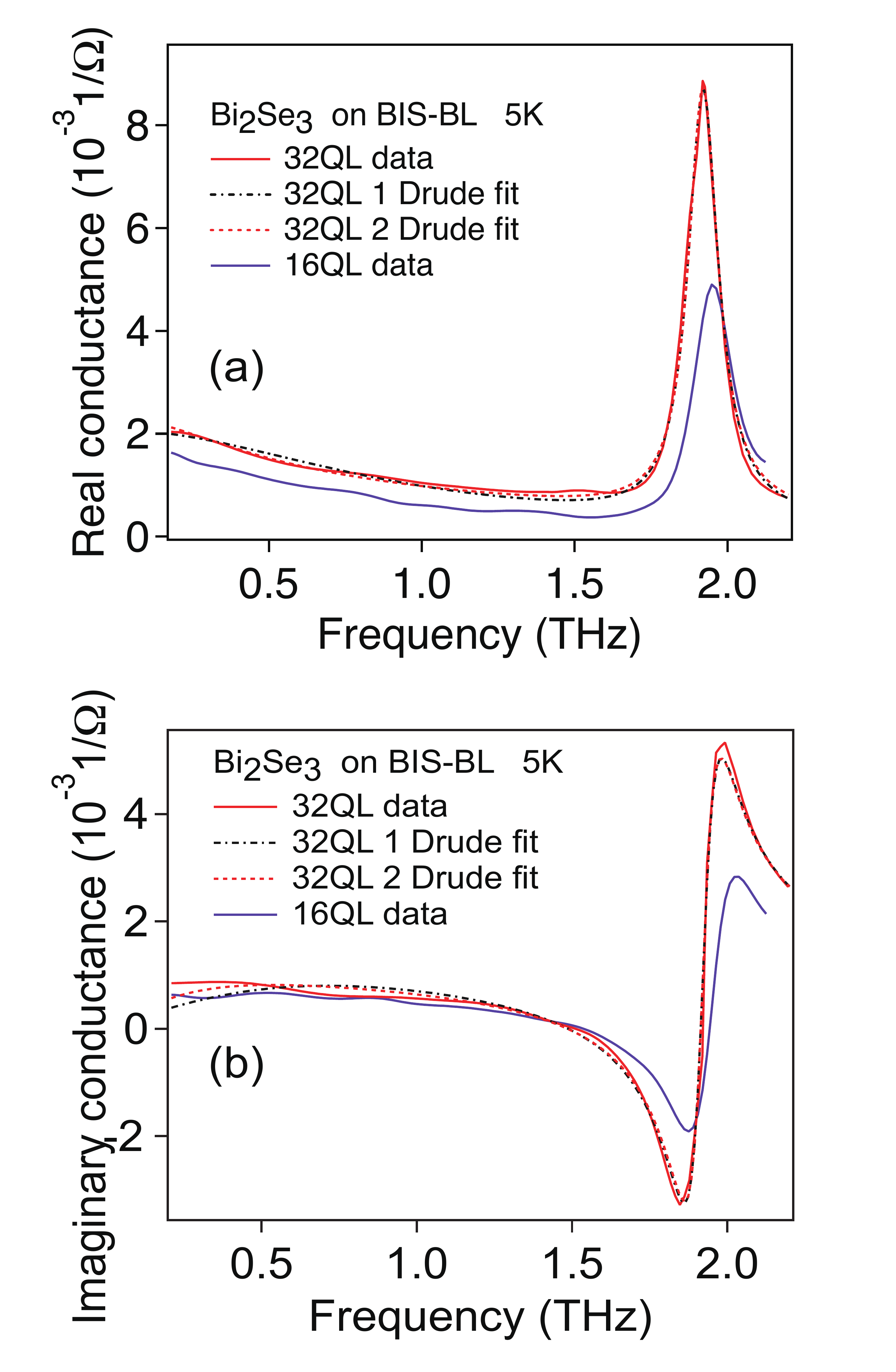}
\caption{(Color online) (a) Real and (b) Imaginary conductance of 32QL and 16QL  Bi$_2$Se$_3$  grown on BIS-BL at 5K. }   \label{Fig1}
\end{figure}

 \begin{figure*}[htp]
\includegraphics[width=\textwidth]{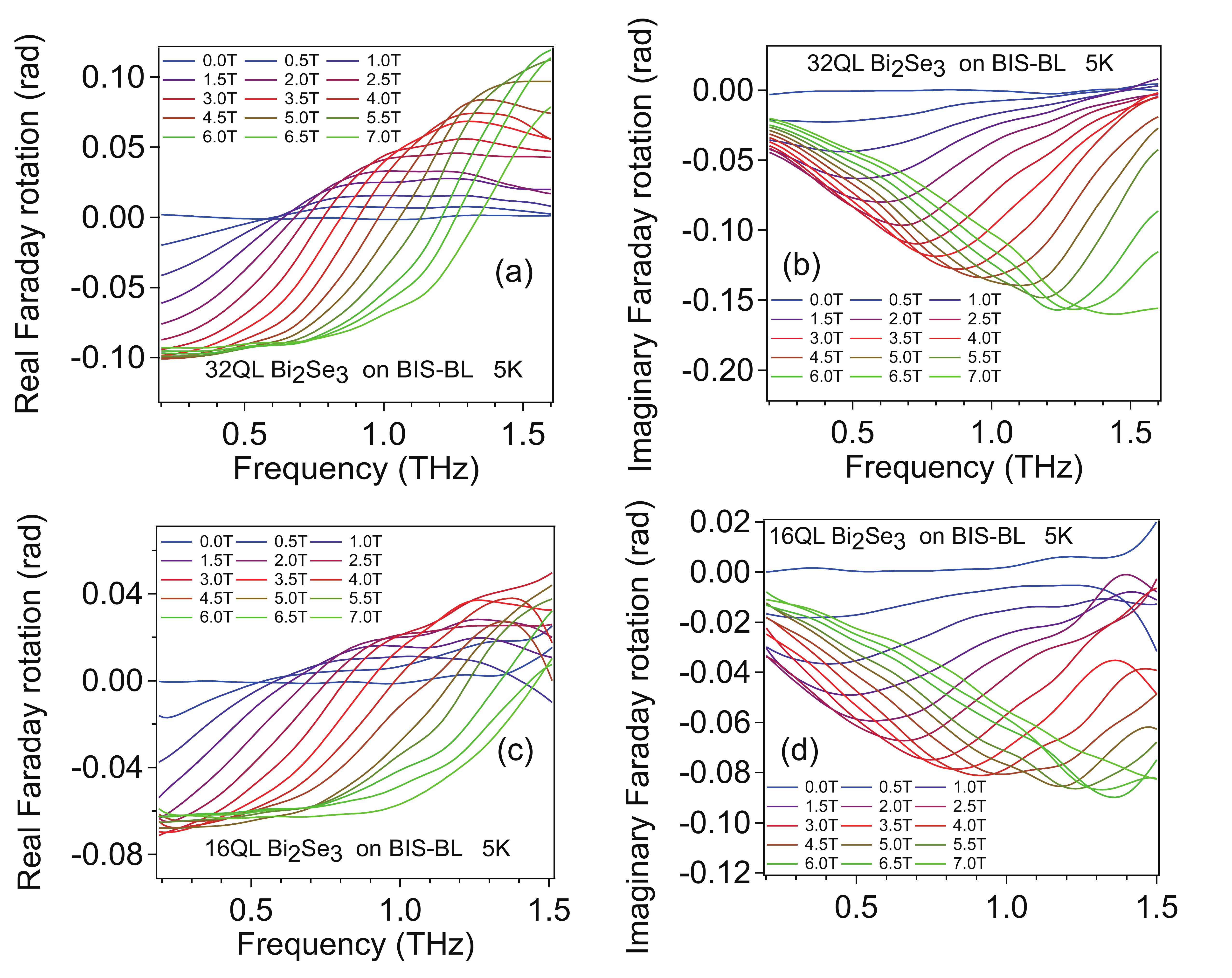}

\caption{(a)Real part and (b) Imaginary part of complex Faraday rotation for 32QL Bi$_2$Se$_3$ grown on BIS-BL layer.  (c)Real part and (d) Imaginary part of complex Faraday rotation for 32QL Bi$_2$Se$_3$.} \label{Fig2}
\end{figure*}

\textbf{Results and Discussion}
 
Here we report time-domain THz spectroscopy measurement on Bi$_2$Se$_3$ thin films with 100 nm Se capping layer and (Bi$_{1-x}$In$_{x}$)$_2$Se$_3$ buffer layer (BIS-BL) \cite{KoiralaNanoLetters15}.  These films have been exposed in air for 6 months. By using cyclotron resonance measurement, we confirm that these films are still bulk-insulating. We attribute the conducting channels to two different topological surface states (TSSs) on the top and bottom of the films. By using magneto-terahertz spectroscopy, we observe that the cyclotron resonance sharpens at higher magnetic field, which reveal with long-range electron impurity scattering on the surfaces.

 \begin{figure*}[htp]
\includegraphics[width=\textwidth]{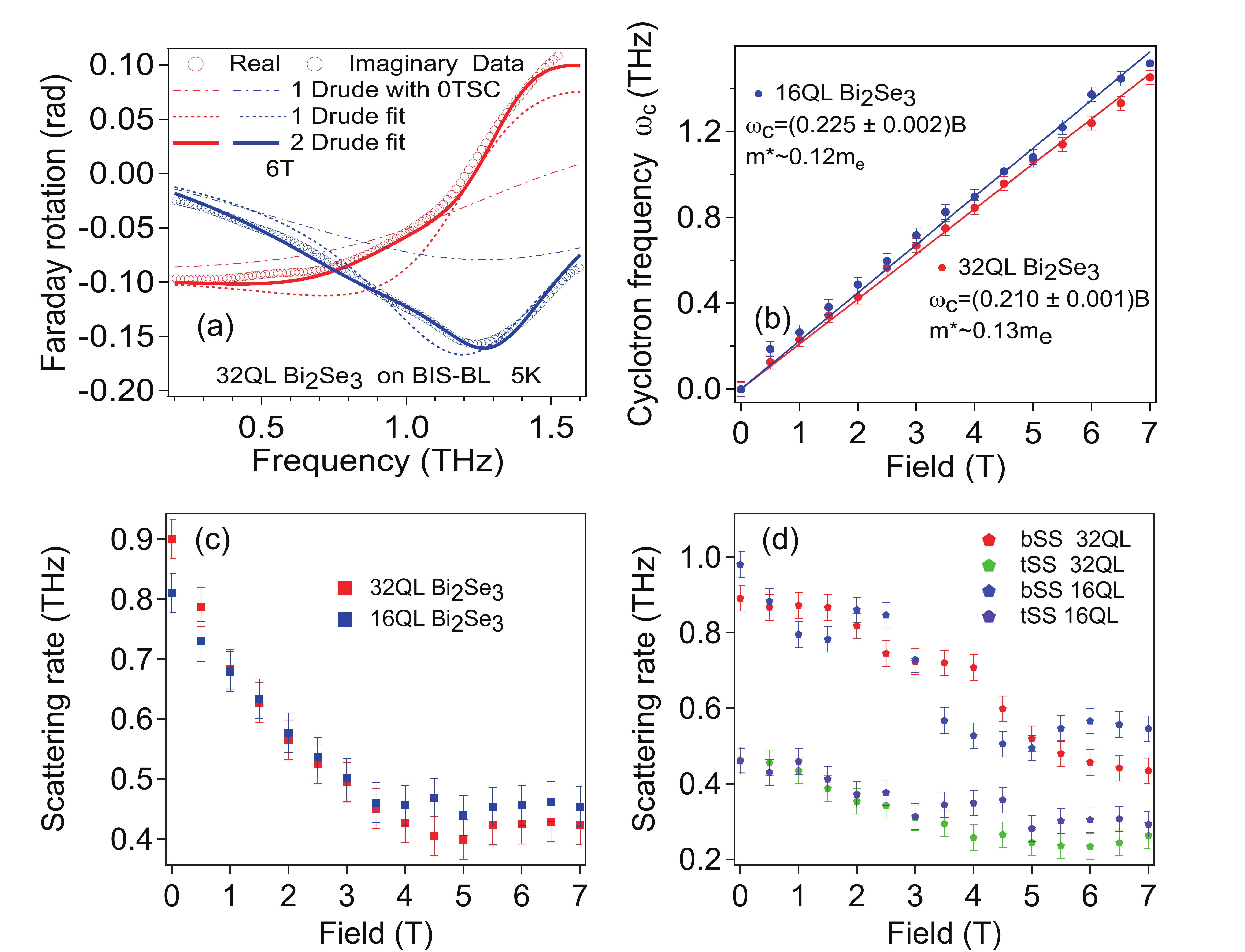}

\caption{(a) A representative fit quality for 1 Drude fit (dashed line) and 2 Drude fit (solid line) and a calculation by using zero-field scattering rate (0TSC) of 32QL sample. (b) Cyclotron frequency VS field from 1 Drude fit. Solid line is linear fit. Drude scattering rate VS field from (c)1 Drude fit (d) 2 Drude fit }.
 \label{Fig3}
\end{figure*}

Fig.\ref{Fig1}(a)(b) shows the real and imaginary conductance for 32QL and 16QL  Bi$_2$Se$_3$  grown on BIS-BL at 5 K. The spectral weight is lower compared with the normal Bi$_2$Se$_3$ directly grown on sapphire or Si \cite{ValdesAguilarPRL12}, even lower than the bulk-insulating Cu$_{0.02}$Bi$_2$Se$_3$ \cite{WuPRL15}. This indicates the bulk insulating behavior of the samples. The spectra can be fitted by the sum of (i) two Drude terms describing free electron-like motions from top and bottom surfaces respectively, (ii) a Drude-Lorentz term describing phonon oscillations, and (iii) a constant term counting for the background above the measured spectral range, as shown below.

\small
\begin{equation}
 G(\omega)=   \epsilon_0 d \left(\sum_{i=1}^2 -\frac{\omega^{2}_{pDi}}{i\omega-\Gamma_{Di}}-\frac{i\omega\omega^{2}_{pDL}}{\omega^{2}_{DL}-\omega^{2}-i\omega\Gamma_{DL}}-i\left(\epsilon_{\infty}-1\right)\omega \right)
\end{equation}
\label{Eqa1}
\normalsize

Here $\Gamma_{Di}$ and $\omega_{pDi}$ are the scattering rates and plasma frequencies of the $i$-th Drude term. $\Gamma_{DL}$ and $\omega_{pDL}$ are the scattering rate and plasma frequency of the Drude-Lorentz term. $d$ is the film thickness. $\epsilon_\infty$ is a constant named lattice polarizability. In the real part of the conductance $G_{D1}$, the total spectral weight ($\omega_{pD}^2 d$) can be extracted from the integral of each feature with a coefficient $\frac{\pi\epsilon_0}{2}$. This gives the ratio of the carrier density and the effective transport mass.


\small
\begin{equation}
\frac{2}{\pi\epsilon_{0}}\int G_{D1} d \omega = \sum_{i=1}^2 \omega_{pDi}^{2} d =\sum_{i=1}^2\frac{n_{2Di}e^{2}}{m_{i}^{*}\epsilon_{0}}
\label{Eqa2}
\end{equation}

\normalsize

In the second equality, $n_{2Di}$ and $m^*_i$ are the carrier densities and the effective transport masses of the massless Dirac fermion. $m^*$ is defined as $m^*=\hbar k_F/v_F.$ Consider a quadratic dispersion model for the TSS, $E=Ak_F+Bk_F^2$, the coefficients $A$ and $B$ can be obtained from photo-emission \cite{WuPRL15}. Then the last term in Eq.\ref{Eqa2} can be expressed by $k_{F}$ only.
For \textit{one} surface state, the relation between spectral weight and $k_F$ is as following. Note that the right side of following expression is half of the value in the Bi$_2$Se$_3$ when we assume two nominally same TSSs were observed in Bi$_2$Se$_3$. 

\small
\begin{equation}
 \omega_{pDi}^{2} d =  \frac{k_F ( A + 2B k_F)e^2 }{ 4 \pi \hbar^2 \epsilon_0}
 \label{Eqa3}
\end{equation}

\normalsize

\noindent At first, a single conduction channel originated from two identical TSSs is assumed. Using Eq.\ref{Eqa2} and \ref{Eqa3}, the $n_{2D}$ and m$^{*}$ can be determined. The fitting results of the 32QL sample gives a \textit{total} sheet carrier density $n_{2D}$ $\sim$ 4.8 $\times$10$^{12}$/cm$^{2}$, effective transport mass $m^{*}\sim$ 0.135 ($\pm$0.005)  $m_{e}$, Fermi energy E$_F \sim$ 140 ($\pm$10) meV, and phonon frequency $\omega_{ph}\sim 1.92$ THz.
16QL sample has lower spectral weight than 32QL sample. The same analysis gives a \textit{total} sheet carrier density $n_{2D}$ $\sim$ 3.8 $\times$10$^{12}$/cm$^{2}$, $m^{*}\sim$ 0.12 ($\pm$0.005) $m_{e}$, E$_F \sim$ 120 ($\pm$8) meV\, and phonon frequency $\omega_{ph}\sim 1.95$ THz. The estimation of the chemical potential is consistent with an insulating bulk even after exposure in air for 6 months \cite{KoiralaNanoLetters15}.

We performed Faraday rotation measurements on these samples to measure the cyclotron resonance mass and spectral weight. Data are shown in Fig.\ref{Fig2}(a)(b)(c)(d). Similar to Cu$_{0.02}$Bi$_2$Se$_3$, cyclotron resonates are observed in both 32QL and 16QL samples.  We use the following model to fit the conductance spectra under an external magnetic field. Here we assume two equal contribution from top and bottom surface states:
\small 
\begin{equation}
G_{\pm}  = -i\epsilon_{0}\omega d  \left(\sum_{i=1}^2 \frac{\omega^{2}_{pDi}}{-\omega^{2}-i\Gamma_{Di}\omega\mp\omega_{c}\omega} +  \frac{\omega^{2}_{pDL}}{\omega^{2}_{DL}-\omega^{2}-i\omega\Gamma_{DL}} +\left(\epsilon_{\infty}-1\right) \right)
\label{Eqa4} 
\end{equation}
\normalsize

\noindent Here the $\pm$ sign corresponds to the right/left-hand circularly polarized light, respectively.  $\omega_{c}$ is the CR frequency. The Faraday rotation can be  expressed by $\mathrm{tan}(\theta_{F})=-i (t_{+}-t_{-})/(t_{+}+t_{-}) $. Here we let $\omega_c$ and $\Gamma_D$ vary with field. We find that the fit result is not good as shown in Fig. \ref{Fig3}(a). Let us further address the 1 Drude model analysis. As we can see in Fig.\ref{Fig3}(a), the fit quality of 1 Drude term is not as good as Cu$_{0.02}$Bi$_2$Se$_3$ case. Bi$_2$Se$_3$ is grown (Bi$_{1-x}$In$_{x}$)$_2$Se$_3$ and $\leq$ 0.2$\%$ In is diffused into the bottom Bi$_2$Se$_3$ few layers. Even though 0.2$\%$ is far below the topological phase transition threshold\cite{WuNatPhys13}, it is natural to treat top surface states (tSS) and bottom surface states (bSS) differently. We remove the equal contribution from both surface states and treat them separately as equation \ref{Eqa4}. The fit quality improve significantly as shown in Fig.\ref{Fig3}(a). We can also see two Drude terms fit zero-field conductance better in Fig.\ref{Fig1}. Then we can extract carrier density of each surface state by the relation  $\omega_{pDi}^{2} d =\frac{n_{2Di}e^{2}}{m_{i}^{*}\epsilon_{0}}$. The parameters for 2 Drude fits are showed in table\ref{tab1}. Indium causes more scattering and therefore the channel with larger scattering rate is identified as bottom surface state. As shown in previous works, the bottom surface has more defects, which lead to higher carrier density \cite{ValdesAguilarPRL12,KoiralaNanoLetters15}. Therefore, the large scattering rate and the large carrier density are consistent with each other for the bottom surface. We also performed Faraday rotation measurements on bare buffer layer and no rotation angle is observed. So the cyclotron resonance feature comes from the Bi$_2$Se$_3$ sample only.


\begin{table}[h!]
\begin{tabular}{ |p{1.6cm}||p{1.4cm}|p{1.1cm}|p{1.6cm}| p{2cm}|}
 \hline
 \hline
  & n$_{2D}$ (10$^{12}$/cm$^2$) & m$^{*}$ (m$_e$) & $\Gamma$(7T) (THz) & $\mu$(7T) (cm$^2$/Vs)\\
 \hline
 tSS 32QL & 2.2 & 0.13 & 0.24  & 8980\\
 bSS 32QL & 2.7 & 0.14 & 0.55  & 3640\\
 tSS 16QL & 1.3 & 0.11 &  0.29 & 8780\\
 bSS 16QL & 2.0 & 0.12 &  0.43 & 5420\\
 \hline
\end{tabular}  
\caption{Fit parameters for two-Drude fit}
\label{tab1}
\end{table}

Now let us address the decreasing scattering rate VS field  as shown in Fig. \ref{Fig3}(c-d). Short-range charge impurity scattering rate leads to $\sqrt B$ increasing scattering rate, as predicted and observed in graphe\red{n}e\cite{Anod2, OrlitaPRL2008, JiangPRL2007}. Electron-phonon scattering rate causes broadening of cyclotron resonance and saturation of scattering rate above phonon energy\cite{WuPRL15}. Long-range charge impurity scattering rate  leads to decreasing scattering rate VS field\cite{yangprb2010}. It was not the dominant mechanism in graphene and therefore was never observed. Similar trend were reported in InSe quantum well and was interpreted as scattering over Gaussian potential larger or equal to the cyclotron orbit\cite{InSe}. The saturation of scattering rate above 4T in Fig. \ref{Fig3}(c-d) is due to electron-phonon scattering as cyclotron frequency reaches $\sim$ 0.7 THz at 4 T in both samples which corresponds to either Kohn anomaly of surface $\beta$ phonon or highest acoustic phonon frequency\cite{WuPRL15}. In the two-Drude fit, one channel has stronger field dependence and is identified as bottom surface state. The reason is charge impurities most likely are created on the bottom surface state due to charge imhomogenity. Top surface state has less effect due to screening.  This kind of long-range charge impurity scattering was further corroborated by the fact no Shubnikov de-Has (SdH) oscillation was observed on these sample while CR is very robust\cite{KoiralaNanoLetters15}. SdH is limited by quantum life-time which is sensitive to both small-angle and large-angle scatting while CR is sensitive to transport life-time which is limited by back scattering. The low quantum life-time in these samples indicates significant small-angle scattering rate coming from long-range coulomb potential. Similar trend was observed in 2DEG system\cite{GaskaAPL1998} and theoretically investigated in literature\cite{AdamPRB2012}. 

To conclude, we observe sharp cyclotron resonance in bulk-insulating Bi$_2$Se$_3$  thin films. The field dependence reveals a decreasing of the scattering rate with increasing magnetic fields. We attribute the origin to impurity scattering due to the buffer layer.  This work shows that cyclotron resonance is a powerful tool to study many-body interactions in topological materials. On the practical side, this work establishes that a 100 nm thick Se capping layer is simple, but very useful for keeping the topological insulator thin films bulk-insulating. The evidence for the long-range impurity scattering at the bottom interface due to buffer layers shreds light on film growth to further decreases the carrier concentration and enhances mobility for future low-dissipation spintronic devices.  \\

\textbf{Acknowledgement} 

This project is supported from the ARO under the Grant W911NF2020166. The development of the THz polarity is supported by the ARO YIP award under the Grant W911NF1910342. The acquisition of the laser for the THz system is  support from a seed grant at National Science Foundation supported University of Pennsylvania Materials Research Science and Engineering Center (MRSEC)(DMR-1720530). X.H. is also partially supported by the Gordon and Betty Moore Foundation’s EPiQS Initiative, Grant GBMF9212 to L.W .  L.W. also acknowledges partial summer support from the NSF EArly-concept Grants for Exploratory Research (EAGER) grant (Grant No. DMR2132591). \\

\textbf{Data Availability Statement}

The data that support the findings of this study are all present in this paper. Further data are available from the corresponding author upon reasonable request.

\bibliography{TopoIns}

\end{document}